\documentclass[12pt]{article}
\usepackage[utf8]{inputenc}
\usepackage[english]{babel}
\usepackage{amsmath}
\usepackage{amsfonts}
\usepackage{amsthm}
\usepackage{amssymb}
\usepackage{stmaryrd}
\usepackage{url}
\usepackage{bbold}
\usepackage[T1]{fontenc}
\usepackage{lmodern}
\usepackage[pdftex]{graphicx}
\usepackage{float}

\begin{document}

\begin{center}
\section*{General Compound Hawkes Processes in Limit Order Books}
{\sc Anatoliy Swishchuk}\footnote{University of Calgary, Calgary, Canada} \footnote{The author wishes to thank NSERC for continuing support}\\
%{\sc Robert Elliott}\footnote{University of Calgary, Calgary, Canada} \footnote{The author wishes to thank NSERC for continuing support}\\
%{\sc Jonathan Chavez Casillas}\footnote{University of Calgary, Calgary, Canada}\\
%{\sc Bruno Remillard}\footnote{HEC, Montreal, Quebec, Canada}\\

\end{center}

\hspace{1cm}

{\bf Abstract:} In this paper, we further study various new Hawkes processes, namely, so-called general compound and regime-switching general compound Hawkes processes to model the price processes in the limit order books. We prove Law of Large Numbers (LLN) and Functional Central Limit Theorems (FCLT) for these processes. The latter two FCLTs are applied to limit order books where we use these asymptotic methods to study the link between price volatility and order flow in our two models by studying the diffusion limits of these price processes. The volatilities of price changes are expressed in terms of parameters describing the arrival rates and price changes. 
%We also present some numerical examples.

\hspace{1cm}

{\bf Keywords}: Hawkes processes; compound Hawkes processes; regime-switching compound Hawkes processes; limit order books; diffusion limits; Law of Large Numbers (LLN); Functional Central Limit Theorem (FCLT)

\section{Introduction}

The Hawkes process (HP) named after its creator Alan Hawkes (1971, 1974) \cite{H1,H2}. The HP is a simple point process that has self-exciting property, clustering effect and long memory. The HP is so-called 'self-exciting point process' which means that it is a point process with stochastic intensity that through its dependence on the history of the process captures the temporal and cross sectional dependence of the event arrival process as well as the 'self-exciting' property observed in the empirical analysis. Self-exciting point processes have recently been applied to high frequency data on price changes or arrival times of trades, namely, market orders, not arrival of limit orders and cancellations. The HP have been used for many applications such as modelling neural activity, genetics Cartensen (2010)\cite{Ca}, occurrence of crime Louie et al. (2010), bank defaults and earthquakes. 
 
Point processes itself gained a significant amount of attention in statistics during the 1950s and 1960s. Cox (1955)\cite{C} introduced the notion of a doubly stochastic Poisson process (called the Cox process now) and Bartlett (1963)\cite{B} investigated statistical methods for point processes based on their power spectral densities.  Lewis (1964)\cite{Le} formulated a point process model (for computer failure patterns) which was a step in the direction of the HP. A nice introduction to the theory of point processes can be found in Daley et al. (1988)\cite{DVJ}. The first type of point processes  in the context of market microstructure is the autoregressive conditional duration (ACD) model introduced by Engel et al. (1998)\cite{ER}.

The most recent application of HP is financial analysis, in particular, limit order books.  In this paper, we further study various new Hawkes processes, namely, so-called compound and regime-switching compound Hawkes processes to model the price processes in the limit order books. We prove Law of Large Numbers and Functional Central Limit Theorems (FCLT) for these processes. The latter two FCLTs are applied to limit order books where we use these asymptotic methods to study the link between price volatility and order flow in our two models by studying the diffusion limits of these price processes. The volatilities of price changes are expressed in terms of parameters describing the arrival rates and price changes. We also present some numerical examples.
The general compound Hawkes process was first introduced in \cite{S} to model the risk process in insurance. In the paper \cite{SCER} we obtained functional CLTs and LLNs for so-called compound Hawkes process with dependent orders and regime-switching compound Hawkes process.

Bowsher (2007)\cite{Bo} was the first one who applied the HP to financial data modelling. Cartea et al. (2011)\cite{CJR} applied HP to model market order arrivals. Filimonov and Sornette (2012)\cite{FSBM} and Filimonov et al. (2013)\cite{FS} apply the HP to estimate the persentage of price changes caused by endogeneous self-generated activity rather than the exogeneous impact of news or novel information. Bauwens and Hautsch (2009)\cite{BH} uses a 5-D HP to estimate multivariate volatility, between five stocks, based on price intencities.   Hewlett (2006)\cite{H} used the instantaneous jump in intensity caused by the occurence of an event to qualify the market impact of that event, taking into account the cascading effect of secondary events also causing futher events. Hewlett (2006)\cite{H} also used the Hawkes model to derive optimal pricing strategies for market makers and optimal trading strategies for investors given that the rational market makers have the historic trading data. Large (2007)\cite{L} applied a Hawkes model for the purpose of invewtigating market impact with more specific interest of order book resiliency and considered limit and market orders and cancellations on both the buy and the sell sides, and further categorizes these events depending on their level of aggression, resulting in a $10$-dimensional Hawkes process. Other econometric models based on marked point processes with stochastic intensity include so-called autoregressive conditional duration (ACD) models by Engle (2000)\cite{E}, Engle and Large (2001)\cite{ELa}. Engle and Lunde (2003)\cite{ELu} considered univariate processes and autoregressive conditional intensity (ACI) models for them with the intensity depeding on the history of the intensity process. Hasbrouck (1999)\cite{Ha} introduced a multivariate point process to model the different events of an order book but did not parametrize the intensity.  We note, that Br\'{e}maud et al. (1996)\cite{BM} generalized the HP to its nonlinear form. Also, a functional central limit theorem for the nonlinear Hawkes process was obtained in Zhu (2013)\cite{Z}. The 'Hawkes diffusion model' introduced in Ait-Sahalia et al. (2010)\cite{AS} in an attempt to extend previous models of stock prices to include financial contagion. Chavez-Demoulin et al. (2012)\cite{CDMG} used Hawkes processes to model high-frequency financial data. Application of affine point processes to portfolio credit risk may be found in Errais et al. (2010)\cite{EGG}. Some applications of Hawkes processes to financial data are also given in Embrechts et al. (2011)\cite{ELL}. 

Cohen et al (2014)\cite{CE} derived an explicit filter for Markov modulated Hawkes process. Vinkovskaya (2014)\cite{V} considered a regime-switching Hawkes process to model its dependency on the bid-ask spread in limot order books.  Regime-switching models for pricing of European and American options were considered in Buffington et al. (2000)\cite{BE1} and Buffington et al. (2002)\cite{ BE2}, respectively. Semi-Markov process was applied to limit order books in \cite{SV} to model the mid-price. We note, that a level-1 limit order books with time dependent arrival rates $\lambda(t)$ were studied in \cite{CCERS}, including the asymptotic distribution of the price process. 

The paper by Bacry et al. (2015)\cite{BMM} proposes an overview of the recent academic literature devoted to the applications of Hawkes processes in finance. It is a nice survey of applications of Hawkes processes in finance.  In general, the main models in high frequency finance can be divided by univariate models, price models, impact models, order-book models, and some systemic risk models, models accounting for news,  high-dimensional models and clustering with graph models. The book by Cartea et al. (2015)\cite{CJP} developed models for algorithmic trading in the contaxts such as executing large orders, market making, trading pairs or collection of assets, and executing in dark pool. This book also contains link to a website, from which many datasets from several sources can be downloaded, and MATLAB code to assist in experimentation with the data. 

A detailed description of the mathematical theory of Hawkes processes is given in Liniger (2009)\cite{Li}. The paper by Laub et al. (2015)\cite{LTP} provides background, introduces the field and historical developments, and touches upon all major aspects of Hawkes processes. 

%%%%%%%%%%%%%%%%%%%%%%
\section{Definition of Hawkes Processes (HPs)}

In this section we give various definitions and some properties of Hawkes processes, that can be found in the existing literature (see, e.g., \cite {H1,H2}, \cite {ELL},  \cite{ZRA}, to name a few). They include, in particular, one-, D-dimensional and non-linear Hawkes processes.

{\bf Definition 1 (Counting Process)}. A counting process is a stochastic process $N(t), t\geq 0,$ taking integer positive values and satisfying: $N(0)=0,$ is almost surely finite, and is right-continuous step function with increments of size $+1.$

Denote by ${\cal F}^N(t), t\geq 0,$ the history of the arrivals up to time $t,$ that is, ${\cal F}^N(t), t\geq 0,$ is a filtration (an increasing sequence of $\sigma$-algebras).

A counting process $N(t)$ can be interpreted as a cumulative count of the number of arrivals into a system up to the current time $t.$ The counting process can also be characterized by the sequence of random arrival times $(T_1,T_2,...)$ at which the counting process $N(t)$ has jumped. The process defined as these arrival times is called a point process (see \cite{DVJ}).

{\bf Definition 2 (Point Process).} If a sequence of random variables $(T_1,T_2,...),$ taking values in $[0,+\infty),$ has $P(0\leq T_1\leq T_2\leq...)=1,$ and the number of points in a bounded region is almost sure finite, then, $(T_1,T_2,...)$ is called a point process.

{\bf Definition 3 (Conditional Intensity Function).} Consider a counting process $N(t)$ with associated histories ${\cal F}^N(t), t\geq 0.$ If a non-negative function $\lambda(t)$ exists such that 
$$
\lambda(t)=\lim_{h\to 0}\frac{E[N(t+h)-N(t)|{\cal F}^N(t)]}{h},
\eqno{(1)}
$$
then it is called the conditional intensity function of $N(t)$ (see \cite{LTP}). We note, that originally this function was called the hazard function (see \cite{C}).

{\bf Definition 4 (One-dimensional Hawkes Process)}. The one-dimensional Hawkes process (see \cite {H1,H2}) is a point point process $N(t)$ which is characterized by its intensity $\lambda(t)$ with respect to its natural filtration:
$$
\lambda (t)=\lambda+\int_{0}^t\mu(t-s)dN(s),
\eqno{(2)}
$$
where $\lambda>0,$ and the response function $\mu(t)$ is a positive function and satisfies $\int_0^{+\infty}\mu(s)ds<1.$ 

The constant $\lambda$ is called the background intensity and the function $\mu(t)$ is sometimes also called excitation function. We suppose that $\mu(t)\not= 0$ to avoid the trivial case, that is, a homogeneous Poisson process. Thus, the Hawkes process is a non-Markovian extension of the Poisson process.

With respect to definitions of $\lambda(t)$ in (1) and $N(t)$ (2), it follows that
$$
P(N(t+h)-N(t)=m|{\cal F}^N(t))=\left\{
\begin{array}{rcl}
\lambda(t)h+o(h), &&m=1\\
o(h),&&m>1\\
1-\lambda(t)h+o(h),&&m=0.\\
\end{array}
\right.
$$

The interpretation of equation (2) is that the events occur according to an intensity with a background intensity $\lambda$ which increases by $\mu(0)$ at each new event then decays back to the background intensity value according to the function $\mu(t).$ Choosing $\mu(0)>0$ leads to a jolt in the intensity at each new event, and this feature is often called self-exciting feature, in other words, if an arrival causes the conditional intensity function $\lambda(t)$ in (1)-(2) to increase then the process is said to be self-exciting. 

We would like to mention that the conditional intensity function $\lambda(t)$ in (1)-(2) can be associated with the compensator $\Lambda(t)$ of the counting process $N(t),$ that is:
$$
\Lambda(t)=\int_0^t\lambda(s)ds.
\eqno{(3)}
$$

Thus, $\Lambda(t)$ is the unique ${\cal F}^N(t), t\geq 0,$ predictable function, with $\Lambda(0)=0,$ and is non-decreasing, such that 
$$
N(t)=M(t)+\Lambda(t)\quad a.s.,
$$
where $M(t)$ is an ${\cal F}^N(t), t\geq 0,$ local martingale (existence of which is guaranteed by the Doob-Meyer decomposition).

A common choice for the function $\mu(t)$ in (2) is one of exponential decay (see \cite{H1}): 
$$
\mu(t)=\alpha e^{-\beta t},
\eqno{(4)}
$$ 
where parameters $\alpha,\beta>0.$ In this case, the Hawkes process is called the Hakes process with exponentially decaying intensity.

Thus, the equation (2) becomes
$$
\lambda (t)=\lambda+\int_{0}^t\alpha e^{-\beta (t-s)}dN(s),
\eqno{(5)}
$$
We note, that in the case of (4), the process $(N(t),\lambda(t))$ is a continuous-time Markov process, which is not the case for the choice (2). 

With some initial condition $\lambda(0)=\lambda_0,$ the conditional density $\lambda(t)$ in (5) with the exponential decay in (4) satisfies the SDE
$$
d\lambda(t)=\beta(\lambda-\lambda(t))dt+\alpha dN(t), \quad t\geq 0,
$$
which can be solved (using stochastic calculus) as
$$
\lambda (t)=e^{-\beta t}(\lambda_0-\lambda)+\lambda+\int_{0}^t\alpha e^{-\beta (t-s)}dN(s),
$$
which is an extension of (5).
 
Another choice for $\mu(t)$ is a power law function:
$$
\lambda (t)=\lambda+\int_{0}^t\frac{k}{(c+(t-s))^p}dN(s)
\eqno{(6)}
$$
with some positive parameters $c,k,p.$ This power law form for $\lambda(t)$ in (6) was applied in the geological model called Omori's law, and used to predict the rate of aftershocks caused by an earthquake. 

{\bf Definition 5 (D-dimensional Hawkes Process)}. The D-dimensional Hawkes process (see \cite {ELL}) is a point point process $\vec N(t)=(N^i(t))_{i=1}^D$ which is characterized by its intensity vector $\vec\lambda(t)=(\lambda^i(t))_{i=1}^D$ such that:
$$
\lambda^i(t)=\lambda^i+\int_{0}^t\mu^{ij}(t-s)dN^j(s),
\eqno{(7)}
$$
where $\lambda^i>0,$ and  $M(t)=(\mu^{ij}(t))$ is a matrix-valued kernel such that:

1) it is component-wise positive: $(\mu^{ij}(t))\geq 0$ for each $1\leq i,j\leq D;$

2) it is component-wise $L^1$-integrable functions.

In matrix-convolution form, the equation (7) can be written as
$$
\vec\lambda(t)=\vec\lambda+M\ast d\vec N(t),
\eqno{(8)}
$$
where $\vec\lambda=(\lambda^i)_{i=1}^D.$

{\bf Definition 6 (Non-linear Hawkes Process).} The non-linear Hawkes process (see, e.g., \cite{ZRA}) is defined by the intensity function in the following form:
$$
\lambda(t)=h\Big(\lambda+\int_{0}^t\mu(t-s)dN(s)\Big),
\eqno{(9)}
$$
where $h(.)$ is a non-linear function with support in $R^+.$ Typical examples for $h$ are $h(x)={\bf 1}_{x\in R^+}$ and $h(x)=e^x.$

{\bf Remark 1}. Many other generalizations  of Hawkes process have been proposed. They include, in particular, mixed diffusion-Hawkes models \cite{EGG}, Hawkes models with shot noise exogenous events \cite{DZ}, Hakes processes with generation dependent kernels \cite{MZ}, to name a few.

%%%%%%%%%%%%%%%%%%%%%%%%%%%%%
\section{General Compound Hawkes Process (GCHP) and Regime-switching General Compound Hawkes Process (RSGCHP)}

In this section, we define general compound Hawkes process (GCHP) and regime-switching compound Hawkes process (RSCHP). We also consider special cases of GCHP and RSGCHP and their applications in limit order books.

\subsection{General Compound Hawkes Process (GCHP) }

{\bf Definition 7 (General Compound Hawkes Process (GCHP)).} Let $N(t)$ be any one-dimensional Hawkes process defined in above section. Let also $X_n$ be ergodic continuous-time finite state Markov chain, independent of $N(t),$ with space state $X,$ and $a(x)$ be any bounded and continuous function on $X.$ The general compound Hawkes process is defined as
$$
S_t=S_0+\sum_{i=1}^{N(t)}a(X_k).
\eqno{(10)}
$$  

{\bf Remark 2.} In similar way, we can define general compound Hawkes processes for other Hawkes processes, such as $D$-dimensional, Definition 5, or non-linear, Definition 6, ones.

\subsubsection{Special Cases of GCHP and Applications: Limit Order Books}

{\bf 3.1.1. 1. (Fixed Tick, Two-values Price Change, Independent Orders)}. If Instead of Markov chain we take the sequence of i.i.d.r.v. $X_k,$  and $a(x)=x,$ then (10) becomes

$$
S_t=S_0+\sum_{i=1}^{N(t)}X_k.
\eqno{(11)}
$$  
In the case of $\mu(t)=0$ in the Definition 1 of Hawkes process $N(t),$ we simply have compound Poisson process $S_t$  in (11)  with $S_0=0.$ 
So, the name of $S_t$ in (10)-compound Hawkes process. We shall call the process $S_t$ in (11)-{\it compound Hawkes process (CHP)} or {\it compound Hawkes process with independent orders (CHPIO)}.
In the case of Poisson process $N(t)$ ($\mu(t)=0$) this model were used in \cite{CdL} to model limit order books with $X_k=\{-\delta,+\delta\},$ where $\delta$ is the fixed tick size.

{\bf 3.1.1.2. (Fixed Tick,  Two-values Price Change, Dependent Orders)}. Suppose that $X_k\in \{-\delta,+\delta\}$ and $a(x)=x,$ then $S_t$ in (10) becomes
$$
S_t=S_0+\sum_{i=1}^{N(t)}X_k.
\eqno{(12)}
$$

This type of process can be a model for the mid-price in the limit order books, where $\delta$ is the fixed tick size and $N(t)$ is the number of order's arrivals up to moment $t.$ 
We shall call the process $S_t$ in (12)-{\it compound Hawkes process with dependent orders (CHPDO)}. $N(t)$ is the renewal process, then the model (12) were used in \cite{SV} for a semi-Markovian modelling of limit order markets.

{\bf 3.1.1.3. (Non-Fixed Tick, Two-values Price Change, Dependent Orders)}. Suppose that $X_k$ is ergodic continuous-time Markov chain, independent on $N(t),$ with two states, $X=\{1,2\},$ and $N(t)$ is the renewal process. The (10) becomes
$$
S_t=S_0+\sum_{i=1}^{N(t)}a(X_k),
\eqno{(13)}
$$
where $a(X_k)$ takes only two values $a(1)$ and $a(2).$ We shall call the process $S_t$ in (13)-{\it general compound Hawkes process with two-state dependent orders (GCHP2SDO)}. This model was used in \cite{SHCS} for mid-price process in limit order books with non-fixed tick $\delta$ and  two-values price change. Of course, this model is more general then (12), as we can consider one tick spread if we set $a(1)=\delta$ and $a(2)=-\delta.$

{\bf 3.1.1.4. (Non-Fixed Tick, $N$-values Price Change, Dependent Orders)}. Suppose that $X_k$ is ergodic continuous-time Markov chain, independent on $N(t),$ with $n$ states, $X=\{1,2,...,n\},$ and $N(t)$ is the renewal process. The (10) becomes
$$
S_t=S_0+\sum_{i=1}^{N(t)}a(X_k),
\eqno{(14)}
$$
where $a(X_k)$ takes $n$ values $a(1), a(2),...,a(n).$ We shall call the process $S_t$ in (14)-{\it general compound Hawkes process with $n$-state dependent orders (GCHPnSDO)}.
This model was used in \cite{SHCS} for mid-price process in limit order books with non-fixed tick $\delta$ and $n$-values price change. This model is even more general then (14), as we can consider $n=2$ for state space $X$ in (14).

%%%%%%%%%%%%%%%%%%%%%%%%%%
\subsection{Regime-switching General Compound Hawkes Process (RSGCHP)}

Let $Y_t$ be an $N$-state Markov chain, with rate matrix1 $A_t.$ We assume, without loss of generality, that $Y_t$ takes values in the standard basis vectors in $R^N.$ Then, $Y_t$ has the representation
$$
Y_t=Y_0+\int_0^tA_sY_sds+M_t,
\eqno{(15)}
$$ 
for $M_t$ an $R^N$ -valued $P$-martingale (see \cite{BE1} more more details).

{\bf Definition 8 (One-dimensional Regime-switching Hawkes Process (RSHP))}. One-dimensional regime-switching Hawkes Process $N_t$ is a point process which characterized by its intensity $\lambda(t)$ in the following way:
$$
\lambda_t=<\lambda,Y_t>+\int_{0}^t<\mu(t-s),Y_s>dN_s,
\eqno{(16)}
$$
where $<\cdot,\cdot>$ is an inner product and $Y_t$ is defined in b(15).

{\bf Definition 9 (D-dimensional Regime-switching Hawkes Process (DRSHP))}. The D-dimensional regime-switching Hawkes process is a point point process $\vec N_t=(N^i_t)_{i=1}^D$ which is characterized by its intensity vector $\vec\lambda_t=(\lambda^i_t)_{i=1}^D$ such that:
$$
\lambda^i_t=<\lambda^i,Y_t>+\int_{0}^t<\mu^{ij}(t-s),Y_s>dN^j_s,
\eqno{(17)}
$$
where $\lambda^i>0,$ and  $M(t)=(\mu^{ij}(t))$ is a matrix-valued kernel defined in (7), Definition 5, and $Y_t$ is defined in (15).

{\bf Definition 10 (Non-linear Regime-switching Hawkes Process (NLRSHP)).} The non-linear regime-switching Hawkes process is defined by the intensity function in the following form:
$$
\lambda_t=h\Big(<\lambda,Y_t>+\int_{0}^t<\mu(t-s),Y_s>dN_s\Big),
\eqno{(18)}
$$
where $h(\cdot)$ is a non-linear function with support in $R^+,$ and $Y_t$ is defined in (15).

{\bf Definition 11 (Regime-switching General Compound Hawkes Process (RSGHP)).} 
Let $N_t$ be any one-dimensional regime-switching Hawkes process (RSHP) defined in (16), Definition 8. Let also $X_n$ be ergodic continuous-time finite state Markov chain, independent of $N_t,$ with space state $X,$ and $a(x)$ be any bounded and continuous function on $X.$ The regime-switching general compound Hawkes process is defined as
$$
S_t=S_0+\sum_{i=1}^{N_t}a(X_k),
\eqno{(19)}
$$  
where $N_t$ is defined in (16), Definition 8.

\subsubsection{Special Cases of RSGCHP and Applications: Limit Order Books}

{\bf 3.2.1.1. (Fixed Tick, Two-values Price Change, Independent Orders)}. If Instead of Markov chain we take the sequence of i.i.d.r.v. $X_k,$ then (19) becomes

$$
S_t=S_0+\sum_{i=1}^{N_t}X_k,
\eqno{(19.1)}
$$  
where $N_t$ is RSHP defined in (16).
In the case of $\mu(t)=0$ in the Definition 1 of Hawkes process $N(t),$ we simply have regime-switching compound Poisson process $S_t$  in (19)  with $S_0=0.$ 
So, the name of $S_t$ in (10)-compound Hawkes process. We shall call the process $S_t$ in (11)-{\it regime-switching compound Hawkes process (RSCHP)} or {\it regime-switching compound Hawkes process with independent orders (RSCHPIO)}.

{\bf 3.2.1.2. (Fixed Tick,  Two-values Price Change, Dependent Orders)}. Suppose that $X_k\in \{-\delta,+\delta\}$ and $a(x)=x,$ then $S_t$ in (19) becomes
$$
S_t=S_0+\sum_{i=1}^{N_t}X_k.
\eqno{(19.2)}
$$

This type of process can be a model for the mid-price in the limit order books, where $\delta$ is the fixed tick size and $N(t)$ is the number of order's arrivals up to moment $t.$ 
We shall call the process $S_t$ in (12)-{\it regime-switching compound Hawkes process with dependent orders (RSCHPDO)}. 

{\bf 3.2.1.3. (Non-Fixed Tick, Two-values Price Change, Dependent Orders)}. Suppose that $X_k$ is ergodic continuous-time Markov chain, independent on $N(t),$ with two states, $X=\{1,2\},$ and $N(t)$ is the renewal process. The (10) becomes
$$
S_t=S_0+\sum_{i=1}^{N_t}a(X_k),
\eqno{(19.3)}
$$
where $a(X_k)$ takes only two values $a(1)$ and $a(2).$ We shall call the process $S_t$ in (13)-{\it regime-switching general compound Hawkes process with two-state dependent orders (RSGCHP2SDO)}.  Of course, this model is more general then (19.2), as we can consider one tick spread if we set $a(1)=\delta$ and $a(2)=-\delta.$

{\bf 3.2.1.4. (Non-Fixed Tick, $N$-values Price Change, Dependent Orders)}. Suppose that $X_k$ is ergodic continuous-time Markov chain, independent on $N(t),$ with $n$ states, $X=\{1,2,...,n\},$ and $N(t)$ is the renewal process. The (10) becomes
$$
S_t=S_0+\sum_{i=1}^{N_t}a(X_k),
\eqno{(19.4)}
$$
where $a(X_k)$ takes $n$ values $a(1), a(2),...,a(n).$ We shall call the process $S_t$ in (14)-{\it regime-switching general compound Hawkes process with $n$-state dependent orders (RSGCHPnSDO)}.
This model is even more general then (19.3), as we can consider $n=2$ for state space $X$ in (14).

{\bf Remark 3.} In similar way, as in Definitions 8-10, we can define regime-switching Hawkes processes with exponential kernel (see (4)) or power law kernel (see (6)).
Accordingly, we can define respective regime-switching general compound Hawkes processes, similar to the one that was defined in (19), Definition 11, if we take $D$-dimensional, Definition 9, or non-linear, Definition 10, regime-switching Hawkes processes $N_t.$

{\bf Remark 4.} Regime-switching Hawkes processes were considered in \cite{CE} (with exponential kernel) and in \cite{V} (multi-dimensional Hawkes process).
Paper \cite{CE} discussed a self-exciting counting process whose parameters depend on a hidden finite-state Markov chain, and the optimal filter and smoother based on observation of the jump process are obtained. Thesis \cite{V} considers a regime-switching multi-dimensional Hawkes process with exponential kernel that reflects changes in the bid-ask spread. 
The statistical properties, such as MLE of its parameters, etc., of this model were studied.

%%%%%%%%%%%%%%%%%%%%%%%%%%%%%
\section{Diffusion Limits and LLNs for Various Hawkes Processes in Limit Order Books}

In this section, we consider LLNs and diffusion limits for various Hawkes processes, defined in section 3, in the limit order books. 
In the limit order books, high-frequency and algorithmic trading, order arrivals and cancellations are very frequent and occur at the millisecond time scale (see, e.g., \cite{CdL}, \cite{CJP}). Meanwhile, in many applications, such as order execution, one is interested in the dynamics of order flow over a large time scale, typically tens of seconds or minutes. It means that we can use asymptotic methods to study the link between price volatility and order flow in our model by studying the diffusion limit of the price process. Here, we prove a functional central limit theorems for the price processes and express the volatilities of price changes in terms of parameters describing the arrival rates and price changes. We also prove LLNs for these various CHPs. In the first section we recap the results from \cite{SCER} for completeness.
We note, that a level-1 limit order books with time dependent arrival rates $\lambda(t)$ were studied in \cite{CCERS}, including the asymptotic distribution of the price process. 

%%%%%%%%%%%%%%%%%%%%%%%%%
\subsection{Diffusion Limit and LLN for Compound Hawkes Process with Dependent Orders (CHPDO) in Limit Order Books}

We consider here the mid-price process $S_t$ (GCHP) which was defined in (12), namely, :
$$
S_t=S_0+\sum_{i=1}^{N(t)}X_k,
\eqno{(20)}
$$
where $X_k\in \{-\delta,+\delta\}$ is continuous-time 2-state Markov chain,  $\delta$ is the fixed tick size, and $N(t)$ is the number of price changes up to moment $t,$ described by one-dimensional Hawkes process defined in (2), Definition 4.  It means that we have the case with fixed tick,  two-values price change and dependent orders.

{\bf Theorem 1 (Diffusion Limit for CHPDO).} Let $X_k$ be an ergodic Markov chain with two states $\{-\delta,+\delta\}$ and with ergodic probabilities $(\pi^*,1-\pi^*).$ Let also $S_t$ is defined in (20). Then
$$
\frac{S_{nt}-N(nt)\cdot s^*}{\sqrt{n}}\to_{n\to+\infty}\sigma\sqrt{\lambda/(1-\hat\mu)}W(t),
\eqno{(21)}
$$
where $W(t)$ is a standard Wiener process, 
$$
0<\hat\mu:=\int_0^{+\infty}\mu(s)ds<1\quad and\quad \int_0^{+\infty}\mu(s)sds<+\infty,
\eqno{(22)}
$$
$$
s^*:=\delta(2\pi^*-1)\quad and\quad \sigma^2:=4\delta^2\Big(\frac{1-p'+\pi^*(p'-p)}{(p+p'-2)^2}-\pi^*(1-\pi^*)\Big),
\eqno{(23)}
$$
and $(p,p')$ are transition probabilities of Markov chain $X_k.$ We note that $\lambda$ and $\mu(t)$ are defined in (2).

{\bf Proof.} From (20) it follows that
$$
S_{nt}=S_0+\sum_{i=1}^{N(nt)}X_k,
\eqno{(24)}
$$
and
$$
S_{nt}=S_0+\sum_{i=1}^{N(nt)}(X_k-s^*)+N(nt)s^*.
$$
Therefore,
$$
\frac{S_{nt}-N(nt)s^*}{\sqrt{n}}=\frac{S_0+\sum_{i=1}^{N(nt)}(X_k-s^*)}{\sqrt{n}}.
\eqno{(25)}
$$
As long as $\frac{S_0}{\sqrt{n}}\to_{n\to+\infty}0,$ we have to find the limit for 
$$\frac{\sum_{i=1}^{N(nt)}(X_k-s^*)}{\sqrt{n}}$$ 
when $n\to+\infty.$

Consider the following sums
$$
R_n:=\sum_{k=1}^{n}(X_k-s^*)
\eqno{(26)}
$$
and
$$
U_n(t):=n^{-1/2}[(1-(nt-\lfloor nt\rfloor))R_{\lfloor nt\rfloor)}+(nt-\lfloor nt\rfloor))R_{\lfloor nt\rfloor)+1}],
\eqno{(27)}
$$
where $\lfloor\cdot\rfloor$ is the floor function.

Following the martingale method from \cite{SV}, we have the following weak convergence in the Skorokhod topology (see \cite{S}):
$$
U_n(t)\to_{n\to+\infty}\sigma W(t),
\eqno{(28)}
$$
where $\sigma$ is defined in (23).

We note, that w.r.t LLN for Hawkes process $N(t)$ (see, e.g., \cite{DVJ}) we have:
$$
\frac{N(t)}{t}\to_{t\to+\infty}\frac{\lambda}{1-\hat\mu},
$$
or
$$
\frac{N(nt)}{n}\to_{n\to+\infty}\frac{t\lambda}{1-\hat\mu},
\eqno{(29)}
$$
where $\hat\mu$ is defined in (22).

Using change of time in (28), $t\to N(nt)/n,$ we can find from (28) and (29):
$$
U_n(N(nt)/n)\to_{n\to+\infty}\sigma W\Big(t\lambda/(1-\hat\mu)\Big),
$$
or
$$
U_n(N(nt)/n)\to_{n\to+\infty}\sigma\sqrt{\lambda/(1-\hat\mu)}W(t),
\eqno{(30)}
$$
The result (21) now follows from (25)-(30).

{\bf Lemma 1 (LLN for CHPDO)}. The process $S_{nt}$ in (24) satisfies the following weak convergence in the Skorokhod topology (see \cite{S}):
$$
\frac{S_{nt}}{n}\to_{n\to+\infty}s^*\cdot\frac{\lambda}{1-\hat\mu}t,
\eqno{(31)}
$$
where $s^*$ and $\hat\mu$ are defined in (23) and (22), respectively.

{\bf Proof.} From (24) we have 
$$
S_{nt}/n=S_0/n+\sum_{i=1}^{N(nt)}X_k/n.
\eqno{(32)}
$$
The first term goes to zero when $n\to+\infty.$ 
From the other side, w.r.t. the strong LLN for Markov chains (see, e.g., \cite{N})
$$
\frac{1}{n}\sum_{k=1}^{n}X_k\to_{n\to+\infty} s^*,
\eqno{(33)}
$$
where $s^*$ is defined in (23).

Finally, taking into account (29) and (33), we obtain:
$$
\sum_{i=1}^{N(nt)}X_k/n=\frac{N(nt)}{n}\frac{1}{N(nt)}\sum_{i=1}^{N(nt)}X_k\to_{n\to+\infty}s^*\frac{\lambda}{1-\hat\mu}t,
$$
and the result in (31) follows.

%%%%%%%%%%%%%%%%%%
 \subsection{Diffusion Limit and LLN for RSCHP with Dependent Orders (RSCHPDO) in Limit Order Books}
 
We consider here the mid-price process $S_t$ (RSGCHP) in the form 
$$
S_t=S_0+\sum_{i=1}^{N_t}X_k,
\eqno{(34)}
$$
where $X_k\in \{-\delta,+\delta\},$ $\delta$ is the fixed tick size, and $N_t$ is the number of price changes up to moment $t,$ described by one-dimensional regime-switching Hawkes process defined as (compare with (16), Definition 8):
$$
\lambda_t=<\lambda,Y_t>+\int_{0}^t\mu(t-s)dN_s.
\eqno{(35)}
$$
Here, we would like to relax the model for one-dimensional regime-switching Hawkes process considering only the case of switching the parameter $\lambda,$ background intensity, in (25), which is reasonable from limit order books' view. For example, we can consider three-state Markov chain $Y_t\in\{1,2,3\}$ and interpret $<\lambda,Y_t>$ as intensity for limit order, say $\lambda_1,$ market order, say $\lambda_2,$ and cancellation order, say $\lambda_3,$ respectively. Of course, more general case (16) can be considered as well, where $<\mu(t),Y_t>,$ excitation function, e.g., can also takes three values, according to limit order, market order and cancellation order, respectively.

{\bf Theorem 2 (Diffusion Limit for RSGCHP)}. Let $X_k$ be an ergodic Markov chain with two states $\{-\delta,+\delta\}$ and with ergodic probabilities $(\pi^*,1-\pi^*).$ Let also $S_t$ is defined in (34) with $\lambda_t$ in (35). We also consider $Y_t$ to be ergodic Markov chain with ergodic probabilities $(p_1^*,p_2^*,...,p_N^*).$ Then
$$
\frac{S_{nt}-N_{nt}\cdot s^*}{\sqrt{n}}\to_{n\to+\infty}\sigma\sqrt{\hat\lambda/(1-\hat\mu)}W(t),
\eqno{(36)}
$$
where $W(t)$ is a standard Wiener process, $s^*$ and $\sigma$ are defined in (23),
$$
\hat\lambda:=\sum_{i=1}^Np_i^*\lambda_i\not=0,\quad \lambda_i:=<\lambda,i>,
\eqno{(37)}
$$
and $\hat\mu$ is defined in (22).

{\bf Proof.} From (34) it follows that
$$
S_{nt}=S_0+\sum_{i=1}^{N_{nt}}X_k,
\eqno{(38)}
$$
and
$$
S_{nt}=S_0+\sum_{i=1}^{N_{nt}}(X_k-s^*)+N_{nt}s^*,
$$
where $N_{nt}$ is RGCHP with regime-switching intensity $\lambda_t$ in (35).
Therefore,
$$
\frac{S_{nt}-N_{nt}s^*}{\sqrt{n}}=\frac{S_0+\sum_{i=1}^{N_{nt}}(X_k-s^*)}{\sqrt{n}}.
\eqno{(39)}
$$
As long as $\frac{S_0}{\sqrt{n}}\to_{n\to+\infty}0,$ we have to find the limit for 
$$\frac{\sum_{i=1}^{N_{nt}}(X_k-s^*)}{\sqrt{n}}$$ 
when $n\to+\infty.$

Consider the following sums, similar to (26) and (27):
$$
R_n:=\sum_{k=1}^{n}(X_k-s^*)
\eqno{(40)}
$$
and
$$
U_n(t):=n^{-1/2}[(1-(nt-\lfloor nt\rfloor))R_{\lfloor nt\rfloor)}+(nt-\lfloor nt\rfloor))R_{\lfloor nt\rfloor)+1}],
\eqno{(41)}
$$
where $\lfloor\cdot\rfloor$ is the floor function.

Following the martingale method from \cite{SV}, we have the following weak convergence in the Skorokhod topology (see \cite{S}):
$$
U_n(t)\to_{n\to+\infty}\sigma W(t),
\eqno{(42)}
$$
where $\sigma$ is defined in (23).

We note, that w.r.t LLN for Hawkes process $N_t$ in (34)  with regime-switching intensity $\lambda_t$ in (35) we have (see \cite{KS} for more details):
$$
\frac{N_{t}}{t}\to_{t\to+\infty}\frac{\hat\lambda}{1-\hat\mu},
$$
or
$$
\frac{N_{nt}}{n}\to_{n\to+\infty}\frac{t\hat\lambda}{1-\hat\mu},
\eqno{(43)}
$$
where $\hat\mu$ is defined in (22) and $\hat\lambda$ in (37).

Using change of time in (43), $t\to N_{nt}/n,$ we can find from (42) and (43):
$$
U_n(N_{nt}/n)\to_{n\to+\infty}\sigma W\Big(t\hat\lambda/(1-\hat\mu)\Big),
$$
or
$$
U_n(N_{nt}/n)\to_{n\to+\infty}\sigma\sqrt{\hat\lambda/(1-\hat\mu)}W(t),
\eqno{(44)}
$$
The result (36) now follows from (38)-(44).

{\bf Lemma 2 (LLN for RSCHPDO)}. The process $S_{nt}$ in (38) satisfies the following weak convergence in the Skorokhod topology (see \cite{S}):
$$
\frac{S_{nt}}{n}\to_{n\to+\infty}s^*\cdot\frac{\hat\lambda}{1-\hat\mu}t,
\eqno{(45)}
$$
where $s^*,$ $\hat\lambda$ and $\hat\mu$ are defined in (23), (37) and (22), respectively.

{\bf Proof.} From (38) we have 
$$
S_{nt}/n=S_0/n+\sum_{i=1}^{N_{nt}}X_k/n,
$$
where $N_{nt}$ is Hawkes process with regime-switching intensity $\lambda_t$ in (35).

The first term goes to zero when $n\to+\infty.$ 

From the other side, w.r.t. the strong LLN for Markov chains (see, e.g., \cite{N})
$$
\frac{1}{n}\sum_{k=1}^{n}X_k\to_{n\to+\infty} s^*,
$$
where $s^*$ is defined in (23).

Finally, taking into account (43) and the last limit, we obtain:
$$
\sum_{i=1}^{N_{nt}}X_k/n=\frac{N_{nt}}{n}\frac{1}{N_{nt}}\sum_{i=1}^{N_{nt}}X_k\to_{n\to+\infty}s^*\frac{\hat\lambda}{1-\hat\mu}t,
$$
and the result in (45) follows.

%%%%%%%%%%%%%%%%%%%
\subsection{Diffusion Limits and LLNs for General Compound Hawkes Process with Two-state Dependent Orders (GCHP2SDO) in Limit Order Books}

We consider here the mid-price process $S_t$ (GCHP) which was defined in (13), namely, :
$$
S_t=S_0+\sum_{i=1}^{N(t)}a(X_k),
\eqno{(46)}
$$
where $X_k\in \{1,2\}:=X$ is continuous-time 2-state Markov chain,  $a(x)$ is continuous and bounded function on $X=\{1,2\},$ and $N(t)$ is the number of price changes up to moment $t,$ described by one-dimensional Hawkes process defined in (2), Definition 4.  It means that we have the case with {\it non-fixed tick},  two-values price change and dependent orders.

{\bf Theorem 3 (Diffusion Limit for GCHP2SDO).} Let $X_k$ be an ergodic Markov chain with two states $\{1,2\}$ and with ergodic probabilities $(\pi^*_1,\pi^*_2).$ Let also $S_t$ is defined in (46). Then
$$
\frac{S_{nt}-N(nt)\cdot a^*}{\sqrt{n}}\to_{n\to+\infty}\sigma^*\sqrt{\lambda/(1-\hat\mu)}W(t),
\eqno{(47)}
$$
where $W(t)$ is a standard Wiener process, 
$$
0<\hat\mu:=\int_0^{+\infty}\mu(s)ds<1\quad and\quad \int_0^{+\infty}\mu(s)sds<+\infty,
\eqno{(48)}
$$
$$
\begin{array}{rcl}
(\sigma^*)^2&:=&\pi^*_1a_1^2+\pi^*_2a_2^2+(\pi^*_1a_1+\pi^*_2a_2)[-2a_1\pi^*_1-2a_2\pi^*_2+(\pi^*_1a_1+\pi^*_2a_2)(\pi^*_1+\pi^*_2)]\\
&+&\frac{(\pi^*_1(1-p)+\pi^*_2(1-p'))(a_1-a_2)^2}{(p+p'-2)^2}\\
&+&2(a_2-a_1)\cdot\Big[\frac{\pi^*_2a_2(1-p')-\pi^*_1a_1(1-p)}{p+p'-2}\\
&+&\frac{(\pi^*_1a_1+\pi^*_2a_2)(\pi^*_1-p\pi^*_1-\pi^*_2+p'\pi^*_2)}{p+p'-2}\Big],\\
a^*&:=&\pi_1^*a(1)+\pi_2^*a(2),\\
\end{array}
\eqno{(49)}
$$
where $(p,p')$ are transition probabilities of Markov chain $X_k.$ We note that $\lambda$ and $\mu(t)$ are defined in (2).

{\bf Proof.} From (46) it follows that
$$
S_{nt}=S_0+\sum_{i=1}^{N(nt)}a(X_k),
\eqno{(50)}
$$
and
$$
S_{nt}=S_0+\sum_{i=1}^{N(nt)}(a(X_k)-a^*)+N(nt)a^*,
$$
where $a^*:=\pi_1^*a(1)+\pi_2^*a(2).$

Therefore,
$$
\frac{S_{nt}-N(nt)a^*}{\sqrt{n}}=\frac{S_0+\sum_{i=1}^{N(nt)}(a(X_k)-a^*)}{\sqrt{n}}.
\eqno{(51)}
$$
As long as $\frac{S_0}{\sqrt{n}}\to_{n\to+\infty}0,$ we have to find the limit for 
$$\frac{\sum_{i=1}^{N(nt)}(a(X_k)-a^*)}{\sqrt{n}}$$ 
when $n\to+\infty.$

Consider the following sums
$$
R^*_n:=\sum_{k=1}^{n}(a(X_k)-a^*)
\eqno{(52)}
$$
and
$$
U^*_n(t):=n^{-1/2}[(1-(nt-\lfloor nt\rfloor))R^*_{\lfloor nt\rfloor)}+(nt-\lfloor nt\rfloor))R^*_{\lfloor nt\rfloor)+1}],
\eqno{(53)}
$$
where $\lfloor\cdot\rfloor$ is the floor function.

Following the martingale method from \cite{SHCS}, we have the following weak convergence in the Skorokhod topology (see \cite{S}):
$$
U^*_n(t)\to_{n\to+\infty}\sigma^* W(t),
\eqno{(54)}
$$
where $\sigma^*$ is defined in (49).

We note again, that w.r.t LLN for Hawkes process $N(t)$ (see, e.g., \cite{DVJ}) we have:
$$
\frac{N(t)}{t}\to_{t\to+\infty}\frac{\lambda}{1-\hat\mu},
$$
or
$$
\frac{N(nt)}{n}\to_{n\to+\infty}\frac{t\lambda}{1-\hat\mu},
\eqno{(55)}
$$
where $\hat\mu$ is defined in (48).

Using change of time in (54), $t\to N(nt)/n,$ we can find from (54) and (55):
$$
U^*_n(N(nt)/n)\to_{n\to+\infty}\sigma^* W\Big(t\lambda/(1-\hat\mu)\Big),
$$
or
$$
U^*_n(N(nt)/n)\to_{n\to+\infty}\sigma^*\sqrt{\lambda/(1-\hat\mu)}W(t),
\eqno{(56)}
$$
The result (47) now follows from (50)-(56).

{\bf Lemma 3 (LLN for GCHP2SDO)}. The process $S_{nt}$ in (46) satisfies the following weak convergence in the Skorokhod topology (see \cite{S}):
$$
\frac{S_{nt}}{n}\to_{n\to+\infty}a^*\cdot\frac{\lambda}{1-\hat\mu}t,
\eqno{(57)}
$$
where $a^*$ and $\hat\mu$ are defined in (49) and (48), respectively.

{\bf Proof.} From (46) we have 
$$
S_{nt}/n=S_0/n+\sum_{i=1}^{N(nt)}a(X_k)/n.
\eqno{(58)}
$$
The first term goes to zero when $n\to+\infty.$ 
From the other side, w.r.t. the strong LLN for Markov chains (see, e.g., \cite{N})
$$
\frac{1}{n}\sum_{k=1}^{n}a(X_k)\to_{n\to+\infty} a^*,
\eqno{(59)}
$$
where $a^*$ is defined in (49).

Finally, taking into account (55) and (59), we obtain:
$$
\sum_{i=1}^{N(nt)}a(X_k)/n=\frac{N(nt)}{n}\frac{1}{N(nt)}\sum_{i=1}^{N(nt)}a(X_k)\to_{n\to+\infty}a^*\frac{\lambda}{1-\hat\mu}t,
$$
and the result in (57) follows.

%%%%%%%%%%%%%%%%%%%
\subsection{Diffusion Limits and LLNs for Regime-switching General Compound Hawkes Process with Two-state Dependent Orders (RSGCHP2SDO) in Limit Order Books}

We consider here the mid-price process $S_t$ (RSGCHP2SDO) in the form 
$$
S_t=S_0+\sum_{i=1}^{N_t}a(X_k),
\eqno{(60)}
$$
where $X_k\in \{1,2\}:=X$ is a continuous-time Markov chain, $a(x)$ is continuous and bounded function on $X=\{1,2\},$ and $N_t$ is the number of price changes up to moment $t,$ described by one-dimensional regime-switching Hawkes process defined as (compare with (16), Definition 8):
$$
\lambda_t=<\lambda,Y_t>+\int_{0}^t\mu(t-s)dN_s.
\eqno{(61)}
$$
We note that $Y_t,$ $N(t)$ and $X_k$ are independent processes.
Here, we would also like to relax the model for one-dimensional regime-switching Hawkes process considering only the case of switching the parameter $\lambda,$ background intensity, in (61), which is reasonable from limit order books' view. For example, we can consider three-state Markov chain $Y_t\in\{1,2,3\}$ and interpret $<\lambda,Y_t>$ as high imbalance, say $\lambda_1,$ normal imbalance, say $\lambda_2,$ and low imbalance, say $\lambda_3,$ respectively. Of course, more general case (16) can be considered here as well, where $<\mu(t),Y_t>,$ excitation function, e.g., can also takes three values, according to high imbalance, normal imbalance and low imbalance, respectively.

{\bf Theorem 4 (Diffusion Limit for RSGCHP2SDO)}. Let $X_k$ be an ergodic Markov chain with two states $\{1,2\}$ and with ergodic probabilities $(\pi^*_1,\pi^*_2).$ Let also $S_t$ is defined in (60) with $\lambda_t$ in (61). We also consider $Y_t$ to be ergodic Markov chain with ergodic probabilities $(p_1^*,p_2^*,...,p_N^*).$ Then
$$
\frac{S_{nt}-N_{nt}\cdot a^*}{\sqrt{n}}\to_{n\to+\infty}\sigma^*\sqrt{\hat\lambda/(1-\hat\mu)}W(t),
\eqno{(62)}
$$
where $W(t)$ is a standard Wiener process, $a^*$ and $\sigma^*$ are defined in (49),
$$
\hat\lambda:=\sum_{i=1}^Np_i^*\lambda_i\not=0,\quad \lambda_i:=<\lambda,i>,
\eqno{(63)}
$$
and $\hat\mu$ is defined in (48).

{\bf Proof.} From (60) it follows that
$$
S_{nt}=S_0+\sum_{i=1}^{N_{nt}}a(X_k),
\eqno{(64)}
$$
and
$$
S_{nt}=S_0+\sum_{i=1}^{N_{nt}}(a(X_k)-a^*)+N_{nt}a^*,
$$
where $N_{nt}$ is RSGCHP with regime-switching intensity $\lambda_t$ in (61).
Therefore,
$$
\frac{S_{nt}-N_{nt}a^*}{\sqrt{n}}=\frac{S_0+\sum_{i=1}^{N_{nt}}(a(X_k)-a^*)}{\sqrt{n}}.
\eqno{(65)}
$$
As long as $\frac{S_0}{\sqrt{n}}\to_{n\to+\infty}0,$ we have to find the limit for 
$$\frac{\sum_{i=1}^{N_{nt}}(a(X_k)-a^*)}{\sqrt{n}}$$ 
when $n\to+\infty.$

Consider the following sums, similar to (26) and (27):
$$
R^*_n:=\sum_{k=1}^{n}(a(X_k)-a^*)
\eqno{(66)}
$$
and
$$
U^*_n(t):=n^{-1/2}[(1-(nt-\lfloor nt\rfloor))R^*_{\lfloor nt\rfloor)}+(nt-\lfloor nt\rfloor))R^*_{\lfloor nt\rfloor)+1}],
\eqno{(67)}
$$
where $\lfloor\cdot\rfloor$ is the floor function.

Following the martingale method from \cite{SHCS}, we have the following weak convergence in the Skorokhod topology (see \cite{S}):
$$
U^*_n(t)\to_{n\to+\infty}\sigma^* W(t),
\eqno{(68)}
$$
where $\sigma^*$ is defined in (49).

We note, that w.r.t LLN for Hawkes process $N_t$ in (34)  with regime-switching intensity $\lambda_t$ in (35) we have (see \cite{KS} for more details):
$$
\frac{N_{t}}{t}\to_{t\to+\infty}\frac{\hat\lambda}{1-\hat\mu},
$$
or
$$
\frac{N_{nt}}{n}\to_{n\to+\infty}\frac{t\hat\lambda}{1-\hat\mu},
\eqno{(69)}
$$
where $\hat\mu$ is defined in (48) and $\hat\lambda$ in (63).

Using change of time in (68), $t\to N_{nt}/n,$ we can find from (68) and (69):
$$
U^*_n(N_{nt}/n)\to_{n\to+\infty}\sigma^* W\Big(t\hat\lambda/(1-\hat\mu)\Big),
$$
or
$$
U^*_n(N_{nt}/n)\to_{n\to+\infty}\sigma^*\sqrt{\hat\lambda/(1-\hat\mu)}W(t),
\eqno{(70)}
$$
The result (62) now follows from (64)-(70).

{\bf Lemma 4 (LLN for RSCHP2SDO)}. The process $S_{nt}$ in (60) satisfies the following weak convergence in the Skorokhod topology (see \cite{S}):
$$
\frac{S_{nt}}{n}\to_{n\to+\infty}a^*\cdot\frac{\hat\lambda}{1-\hat\mu}t,
\eqno{(71)}
$$
where $a^*,$ $\hat\lambda$ and $\hat\mu$ are defined in (49), (63) and (48), respectively.

{\bf Proof.} From (60) we have 
$$
S_{nt}/n=S_0/n+\sum_{i=1}^{N_{nt}}a(X_k)/n,
\eqno{(72)}
$$
where $N_{nt}$ is Hawkes process with regime-switching intensity $\lambda_t$ in (61).

The first term goes to zero when $n\to+\infty.$ 

From the other side, w.r.t. the strong LLN for Markov chains (see, e.g., \cite{N})
$$
\frac{1}{n}\sum_{k=1}^{n}a(X_k)\to_{n\to+\infty} a^*,
\eqno{(73)}
$$
where $a^*$ is defined in (49).

Finally, taking into account (68) and (73), we obtain:
$$
\sum_{i=1}^{N_{nt}}a(X_k)/n=\frac{N_{nt}}{n}\frac{1}{N_{nt}}\sum_{i=1}^{N_{nt}}a(X_k)\to_{n\to+\infty}a^*\frac{\hat\lambda}{1-\hat\mu}t,
$$
and the result in (71) follows.

{\bf Remark 5.} The results of Theorems 3-4 and Lemmas 3-4 are more general than results of Theorems 1-2 and Lemmas 1-2. To see that, it is enough to take in Theorems 3-4 and Lemmas 3-4 $a(1)=\delta,$ $a(2)=-\delta,$ $\pi^*_1=\pi^*,$ $\pi_2^*=1-\pi^*,$ and then we get $a^*:=\pi_1^*a(1)+\pi_2^*a(2)=\pi^*\delta+(1-\pi^*)(-\delta)=\delta(2\pi^*-1)=s^*$ (exactly as in Theorems 1-2 and in Lemmas 1-2). Similarly we can check that $\sigma^*$ in Theorems 3-4 coincides with $\sigma$ in Theorems 1-2: $\sigma^*=\sigma$ (see \cite{SHCS} for detailed calculation).

%%%%%%%%%%%%%%%%%%%%
\subsection{Diffusion Limits and LLNs for General Compound Hawkes Process with $n$-state Dependent Orders (GCHPnSDO) in Limit Order Books}

We consider here the mid-price process $S_t$ (GCHPnSDO) which was defined in (14), namely, :
$$
S_t=S_0+\sum_{i=1}^{N(t)}a(X_k),
\eqno{(74)}
$$
where $X_k\in \{1,2,...n\}:=X$ is continuous-time $n$-state Markov chain,  $a(x)$ is continuous and bounded function on $X=\{1,2,...,n\},$ and $N(t)$ is the number of price changes up to moment $t,$ described by one-dimensional Hawkes process defined in (2), Definition 4.  It means that we have the case with {\it non-fixed tick},  $n$-values price change and dependent orders.

{\bf Theorem 5 (Diffusion Limit for GCHPnSDO).} Let $X_k$ be an ergodic Markov chain with $n$ states $\{1,2,...,n\}$ and with ergodic probabilities $(\pi^*_1,\pi^*_2,...,\pi^*_n).$ Let also $S_t$ is defined in (74). Then
$$
\frac{S_{nt}-N(nt)\cdot\hat a^*}{\sqrt{n}}\to_{n\to+\infty}\hat\sigma^*\sqrt{\lambda/(1-\hat\mu)}W(t),
\eqno{(75)}
$$
where $W(t)$ is a standard Wiener process, 
$$
0<\hat\mu:=\int_0^{+\infty}\mu(s)ds<1\quad and\quad \int_0^{+\infty}\mu(s)sds<+\infty,
\eqno{(76)}
$$
$(\hat\sigma^*)^2:=\sum_{i \in X} \pi^*_iv(i)$
$$
\begin{array}{rcl}
v(i)&=& b(i)^2+\sum_{j\in X}(g(j)-g(i))^2P(i,j)-2b(i)\sum_{j\in\mathcal{S}}(g(j)-g(i))P(i,j),\\
b&=&(b(1),b(2),...,b(n))',\\
b(i):&=&a(X_i)-a^*:=a(i)-a^*, \\
g:&=&(P+\Pi^*-I)^{-1}b,\\
\hat a^*&:=&\sum_{i\in X}\pi^*_ia(X_i),\\
\end{array}
\eqno{(77)}
$$
$P$ is a transition probability matrix for $X_k,$, i.e., $P(i,j)=P(X_{k+1}=j|X_k=i)$. $\Pi^*$ denotes the matrix of stationary distributions of $P$ and $g(j)$ is the jth entry of $g.$

{\bf Proof.} From (74) it follows that
$$
S_{nt}=S_0+\sum_{i=1}^{N(nt)}a(X_k),
\eqno{(78)}
$$
and
$$
S_{nt}=S_0+\sum_{i=1}^{N(nt)}(a(X_k)-\hat a^*)+N(nt)\hat a^*,
$$
where $\hat a^*:=\pi_1^*a(1)+\pi_2^*a(2)+...+\pi^*_na(n)$ (see (77)).

Therefore,
$$
\frac{S_{nt}-N(nt)\hat a^*}{\sqrt{n}}=\frac{S_0+\sum_{i=1}^{N(nt)}(a(X_k)-\hat a^*)}{\sqrt{n}}.
\eqno{(79)}
$$
As long as $\frac{S_0}{\sqrt{n}}\to_{n\to+\infty}0,$ we have to find the limit for 
$$\frac{\sum_{i=1}^{N(nt)}(a(X_k)-\hat a^*)}{\sqrt{n}}$$ 
when $n\to+\infty.$

Consider the following sums
$$
\hat R^*_n:=\sum_{k=1}^{n}(a(X_k)-\hat a^*)
\eqno{(80)}
$$
and
$$
\hat U^*_n(t):=n^{-1/2}[(1-(nt-\lfloor nt\rfloor))\hat R^*_{\lfloor nt\rfloor)}+(nt-\lfloor nt\rfloor))\hat R^*_{\lfloor nt\rfloor)+1}],
\eqno{(81)}
$$
where $\lfloor\cdot\rfloor$ is the floor function.

Following the martingale method from \cite{SHCS}, we have the following weak convergence in the Skorokhod topology (see \cite{S}):
$$
\hat U^*_n(t)\to_{n\to+\infty}\hat\sigma^* W(t),
\eqno{(82)}
$$
where $\hat\sigma^*$ is defined in (77).

We note again, that w.r.t LLN for Hawkes process $N(t)$ (see, e.g., \cite{DVJ}) we have:
$$
\frac{N(t)}{t}\to_{t\to+\infty}\frac{\lambda}{1-\hat\mu},
$$
or
$$
\frac{N(nt)}{n}\to_{n\to+\infty}\frac{t\lambda}{1-\hat\mu},
\eqno{(83)}
$$
where $\hat\mu$ is defined in (76).

Using change of time in (82), $t\to N(nt)/n,$ we can find from (82) and (83):
$$
\hat U^*_n(N(nt)/n)\to_{n\to+\infty}\hat\sigma^* W\Big(t\lambda/(1-\hat\mu)\Big),
$$
or
$$
\hat U^*_n(N(nt)/n)\to_{n\to+\infty}\hat\sigma^*\sqrt{\lambda/(1-\hat\mu)}W(t),
$$
The result (75) now follows from (78)-(83) and the last limit.

{\bf Lemma 5 (LLN for GCHPnSDO)}. The process $S_{nt}$ in (74) satisfies the following weak convergence in the Skorokhod topology (see \cite{S}):
$$
\frac{S_{nt}}{n}\to_{n\to+\infty}\hat a^*\cdot\frac{\lambda}{1-\hat\mu}t,
\eqno{(84)}
$$
where $\hat a^*$ and $\hat\mu$ are defined in (77) and (76), respectively.

{\bf Proof.} From (74) we have 
$$
S_{nt}/n=S_0/n+\sum_{i=1}^{N(nt)}a(X_k)/n.
\eqno{(85)}
$$
The first term goes to zero when $n\to+\infty.$ 
From the other side, w.r.t. the strong LLN for Markov chains (see, e.g., \cite{N})
$$
\frac{1}{n}\sum_{k=1}^{n}a(X_k)\to_{n\to+\infty} \hat a^*,
\eqno{(86)}
$$
where $\hat a^*$ is defined in (77).

Finally, taking into account (83) and (86), we obtain:
$$
\sum_{i=1}^{N(nt)}a(X_k)/n=\frac{N(nt)}{n}\frac{1}{N(nt)}\sum_{i=1}^{N(nt)}a(X_k)\to_{n\to+\infty}\hat a^*\frac{\lambda}{1-\hat\mu}t,
$$
and the result in (84) follows.

%%%%%%%%%%%%%%%%%%%%
\subsection{Diffusion Limits and LLNs for Regime-switching General Compound Hawkes Process with $n$-state Dependent Orders (RSGCHPnSDO) in Limit Order Books}

We consider here the mid-price process $S_t$ (RSGCHPnSDO) in the form 
$$
S_t=S_0+\sum_{i=1}^{N_t}a(X_k),
\eqno{(87)}
$$
where $X_k\in \{1,2,...,n\}:=X$ is a continuous-time $n$-state Markov chain, $a(x)$ is continuous and bounded function on $X=\{1,2,...,n\},$ and $N_t$ is the number of price changes up to moment $t,$ described by one-dimensional regime-switching Hawkes process defined as (compare with (16), Definition 8):
$$
\lambda_t=<\lambda,Y_t>+\int_{0}^t\mu(t-s)dN_s.
\eqno{(88)}
$$
We note that $Y_t,$ $N(t)$ and $X_k$ are independent processes.
Here, we would also like to relax the model for one-dimensional regime-switching Hawkes process considering only the case of switching the parameter $\lambda,$ background intensity, in (88), which is reasonable from limit order books' view. For example, we can consider three-state Markov chain $Y_t\in\{1,2,3\}$ and interpret $<\lambda,Y_t>$ as high imbalance, say $\lambda_1,$ normal imbalance, say $\lambda_2,$ and low imbalance, say $\lambda_3,$ respectively. Of course, more general case (16) can be considered here as well, where $<\mu(t),Y_t>,$ excitation function, e.g., can also takes three values, according to high imbalance, normal imbalance and low imbalance, respectively.

{\bf Theorem 6 (Diffusion Limit for RSGCHPnSDO).} Let $X_k$ be an ergodic Markov chain with $n$ states $\{1,2,...,n\}$ and with ergodic probabilities $(\pi^*_1,\pi^*_2,...,\pi^*_n).$ Let also $S_t$ is defined in (87). We also consider $Y_t$ to be ergodic Markov chain with ergodic probabilities $(p_1^*,p_2^*,...,p_N^*).$ Then
$$
\frac{S_{nt}-N(nt)\cdot\hat a^*}{\sqrt{n}}\to_{n\to+\infty}\hat\sigma^*\sqrt{\hat\lambda/(1-\hat\mu)}W(t),
\eqno{(89)}
$$
where $W(t)$ is a standard Wiener process, 
$$
0<\hat\mu:=\int_0^{+\infty}\mu(s)ds<1\quad and\quad \int_0^{+\infty}\mu(s)sds<+\infty,
\eqno{(90)}
$$
$$
\hat\lambda:=\sum_{i=1}^Np_i^*\lambda_i\not=0,\quad \lambda_i:=<\lambda,i>,
\eqno{(91)}
$$

$(\hat\sigma^*)^2:=\sum_{i \in X} \pi^*_iv(i)$
$$
\begin{array}{rcl}
v(i)&=& b(i)^2+\sum_{j\in X}(g(j)-g(i))^2P(i,j)-2b(i)\sum_{j\in\mathcal{S}}(g(j)-g(i))P(i,j),\\
b&=&(b(1),b(2),...,b(n))',\\
b(i):&=&a(X_i)-a^*:=a(i)-a^*, \\
g:&=&(P+\Pi^*-I)^{-1}b,\\
\hat a^*&:=&\sum_{i\in X}\pi^*_ia(X_i),\\
\end{array}
\eqno{(92)}
$$
$P$ is a transition probability matrix for $X_k,$, i.e., $P(i,j)=P(X_{k+1}=j|X_k=i)$. $\Pi^*$ denotes the matrix of stationary distributions of $P$ and $g(j)$ is the jth entry of $g.$

{\bf Proof.} From (87) it follows that
$$
S_{nt}=S_0+\sum_{i=1}^{N(nt)}a(X_k),
\eqno{(93)}
$$
and
$$
S_{nt}=S_0+\sum_{i=1}^{N(nt)}(a(X_k)-\hat a^*)+N(nt)\hat a^*,
$$
where $\hat a^*:=\pi_1^*a(1)+\pi_2^*a(2)+...+\pi^*_na(n)$ (see (92)).

Therefore,
$$
\frac{S_{nt}-N(nt)\hat a^*}{\sqrt{n}}=\frac{S_0+\sum_{i=1}^{N(nt)}(a(X_k)-\hat a^*)}{\sqrt{n}}.
\eqno{(94)}
$$
As long as $\frac{S_0}{\sqrt{n}}\to_{n\to+\infty}0,$ we have to find the limit for 
$$\frac{\sum_{i=1}^{N(nt)}(a(X_k)-\hat a^*)}{\sqrt{n}}$$ 
when $n\to+\infty.$

Consider the following sums
$$
\hat R^*_n:=\sum_{k=1}^{n}(a(X_k)-\hat a^*)
\eqno{(95)}
$$
and
$$
\hat U^*_n(t):=n^{-1/2}[(1-(nt-\lfloor nt\rfloor))\hat R^*_{\lfloor nt\rfloor)}+(nt-\lfloor nt\rfloor))\hat R^*_{\lfloor nt\rfloor)+1}],
\eqno{(96)}
$$
where $\lfloor\cdot\rfloor$ is the floor function.

Following the martingale method from \cite{SHCS}, we have the following weak convergence in the Skorokhod topology (see \cite{S}):
$$
\hat U^*_n(t)\to_{n\to+\infty}\hat\sigma^* W(t),
\eqno{(97)}
$$
where $\hat\sigma^*$ is defined in (92).

We note, that w.r.t LLN for Hawkes process $N_t$ with regime-switching intensity $\lambda_t$ in (88) we have (see \cite{KS} for more details):
$$
\frac{N_{t}}{t}\to_{t\to+\infty}\frac{\hat\lambda}{1-\hat\mu},
$$
or
$$
\frac{N_{nt}}{n}\to_{n\to+\infty}\frac{t\hat\lambda}{1-\hat\mu},
\eqno{(98)}
$$
where $\hat\mu$ is defined in (90) and $\hat\lambda$ in (91).

Using change of time in (97), $t\to N(nt)/n,$ we can find from (97) and (98):
$$
\hat U^*_n(N(nt)/n)\to_{n\to+\infty}\hat\sigma^* W\Big(t\hat\lambda/(1-\hat\mu)\Big),
$$
or
$$
\hat U^*_n(N(nt)/n)\to_{n\to+\infty}\hat\sigma^*\sqrt{\hat\lambda/(1-\hat\mu)}W(t),
\eqno{(99)}
$$
The result (89) now follows from (93)-(99).

{\bf Lemma 6 (LLN for RSGCHPnSDO)}. The process $S_{nt}$ in (87) satisfies the following weak convergence in the Skorokhod topology (see \cite{S}):
$$
\frac{S_{nt}}{n}\to_{n\to+\infty}\hat a^*\cdot\frac{\hat\lambda}{1-\hat\mu}t,
\eqno{(100)}
$$
where $\hat a^*$ and $\hat\mu$ are defined in (92) and (91), respectively.

{\bf Proof.} From (87) we have 
$$
S_{nt}/n=S_0/n+\sum_{i=1}^{N(nt)}a(X_k)/n.
\eqno{(101)}
$$
The first term goes to zero when $n\to+\infty.$ 
From the other side, w.r.t. the strong LLN for Markov chains (see, e.g., \cite{N})
$$
\frac{1}{n}\sum_{k=1}^{n}a(X_k)\to_{n\to+\infty} \hat a^*,
\eqno{(102)}
$$
where $\hat a^*$ is defined in (92).

Finally, taking into account (98) and (102), we obtain:
$$
\sum_{i=1}^{N(nt)}a(X_k)/n=\frac{N(nt)}{n}\frac{1}{N(nt)}\sum_{i=1}^{N(nt)}a(X_k)\to_{n\to+\infty}\hat a^*\frac{\hat\lambda}{1-\hat\mu}t,
$$
and the result in (100) follows.

{\bf Remark 6.} The results of Theorems 5-6 and Lemmas 5-6, sections 4.5-4.6, are even more general than results of Theorems 3-4 and Lemmas 3-4, sections 4.3.-4.4.
To see that, we can take the state space $X$ containing only two states $X=\{1,2\}$, then we get
$$
\begin{array}{rcl}
 v(1)&=&b(1)^2+P(1,2)(g(2)-g(1))^2-2b(1)P(1,2)(g(2)-g(1)),\\
 v(2)&=&b(2)^2+P(2,1)(g(1)-g(2))^2-2b(2)P(2,1)(g(1)-g(2)),\\
\end{array}
$$
where $b(i):=a(i)-a^*,$ with $P(1,2)=p_1$, $P(2,1)=p_2$, $\vec g=(P+\Pi^*-I)^{-1}(a_1-a^*,a_2-a^*)'$ and $\vec g=(g(1),g(2)).$

%%%%%%%%%%%%%%%%%%%%%
\subsection{Diffusion Limits and LLNs for Non-linear Compound Hawkes Process with $n$-state Dependent Orders (NLCHPnSDO) in Limit Order Books}

In this section, we consider the mid-price process $S_t$ (NLCHPnSDO) in the form 
$$
S_t=S_0+\sum_{i=1}^{N_t}a(X_k),
\eqno{(103)}
$$
where $X_k\in \{1,2,...,n\}:=X$ is a continuous-time $n$-state Markov chain, $a(x)$ is continuous and bounded function on $X=\{1,2,...,n\},$  $N(t)$ is the non-linear Hawkes process (see, e.g., \cite{ZRA}) defined by the intensity function in the following form (see (9), Definition 6):
$$
\lambda(t)=h\Big(\lambda+\int_{0}^t\mu(t-s)dN(s)\Big),
\eqno{(104)}
$$
where $h(.)$ is a non-linear increasing function with support in $R^+,$ $\alpha$-Lipschitz (see \cite{BM}) and such that $\alpha ||h||_{L^1}<1.$ Under the latter conditions, it was proved in \cite{BM} that there exists a unique stationary and ergodic Hawkes process satisfying the dynamics (104). 

{\bf Theorem 7 (Diffusion Limit for NLCHPnSDO)}. Let $X_k$ be an ergodic Markov chain with $n$ states $\{1,2,...,n\}$ and with ergodic probabilities $(\pi^*_1,\pi^*_2,...,\pi^*_n).$ Let also $S_t$ is defined in (103) with non-linear function $h(t)$ satisfying the conditions in (104). Then
$$
\frac{S_{nt}-N(nt)\cdot\hat a^*}{\sqrt{n}}\to_{n\to+\infty}\hat\sigma^*\sqrt{E[N[0,1]]}W(t),
\eqno{(105)}
$$
where $W(t)$ is a standard Wiener process, 
$$
0<\hat\mu:=\int_0^{+\infty}\mu(s)ds<1\quad and\quad \int_0^{+\infty}\mu(s)sds<+\infty,
\eqno{(106)}
$$
$\hat\sigma^*$ is defined in (77), $\hat a^*$ in (92) and $E[N[0,1]]$ is the mean of $N[0,1]$ (the number of points in the interval $[0,1]$) under the stationary and ergodic measure.

{\bf Proof.}  We note, that the result from \cite{BM} implies that, by ergodic theorem,
$$
\frac{N(t)}{t}\to_{t\to+\infty} E[N[0,1]]
$$
or
$$
\frac{N(nt)}{n}\to_{t\to+\infty} tE[N[0,1]].
\eqno{(107)}
$$
We proceed now with similar derivations as in (78)-(82), where we have $N(t)$ as non-linear Hawkes process defined in (104).  Using change of time in (82) we obtain:
$$
\hat U^*_n(N(nt)/n)\to_{n\to+\infty}\hat\sigma^* W\Big(tE[N[0,1]]\Big),
$$
or
$$
\hat U^*_n(N(nt)/n)\to_{n\to+\infty}\hat\sigma^*\sqrt{E[N[0,1]]}W(t),
\eqno{(108)}
$$
The result (105) now follows from (107)-(108).

{\bf Lemma 7 (LLN for NLCHPnSDO)}. The process $S_{nt}$ in (103) satisfies the following weak convergence in the Skorokhod topology:
$$
\frac{S_{nt}}{n}\to_{n\to+\infty}\hat a^*E[N[0,1]]t,
\eqno{(109)}
$$
where $\hat a^*$ is defined in (92) and $E[N[0,1]]$ in (105).

{\bf Proof.} The result (109) follows from (101)-(102), with $N(nt)$ replaced by non-linear Hawkes process $N_{nt},$ and from (107).

%%%%%%%%%%%%%%%%%%%%%%
\subsection{Diffusion Limits and LLNs for Non-linear Regime-switching Compound Hawkes Process with $n$-state Dependent Orders (NLRSCHPnSDO) in Limit Order Books}

Here, we consider the mid-price process $S_t$ (NLRSCHPnSDO) in the form 
$$
S_t=S_0+\sum_{i=1}^{N_t}a(X_k),
\eqno{(109)}
$$
where $X_k\in \{1,2,...,n\}:=X$ is a continuous-time $n$-state Markov chain, $a(x)$ is continuous and bounded function on $X=\{1,2,...,n\},$  $N_t$ is the non-linear regime-switching Hawkes process defined by the intensity function in the following form (compare with (18), Definition 10):
$$
\lambda_t=h\Big(<\lambda,Y_t>+\int_{0}^t\mu(t-s)dN_s\Big),
\eqno{(110)}
$$
where $Y_t$ is $N$-state Markov chain with values in the standard basis vectors in $R^N$ (see 15),  $h(\cdot)$ is a non-linear increasing function with support in $R^+,$ $\alpha$-Lipschitz (see \cite{BM}) and such that $\alpha ||h||_{L^1}<1.$ We note again, that under the latter conditions, there exists a unique stationary and ergodic Hawkes process satisfying the dynamics (110) for every $Y_t=j.$ 
We also define $N_{[0,t]},$ as the number of points in the interval $[0,t]$ for Hawkes process $N_t,$ and $N^i_{[0,t]}$ as the number of points in the interval $[0,t]$ for Hawkes process $N^i_t$ defined by the intensity function in the following way:
$$
\lambda^i_t=h\Big(<\lambda,i>+\int_{0}^t\mu(t-s)dN^i_s\Big),
\eqno{(111)}$$ 

{\bf Theorem 8 (Diffusion Limit for NLRSCHPnSDO)} Let $X_k$ be an ergodic Markov chain with $n$ states $\{1,2,...,n\}$ and with ergodic probabilities $(\pi^*_1,\pi^*_2,...,\pi^*_n).$ Let also $S_t$ is defined in (109). We also consider $Y_t$ to be ergodic Markov chain with ergodic probabilities $(p_1^*,p_2^*,...,p_N^*).$ Then
$$
\frac{S_{nt}-N(nt)\cdot\hat a^*}{\sqrt{n}}\to_{n\to+\infty}\hat\sigma^*\sqrt{\sum_{i=1}^{N}p_i^*E[N^i_{[0,1]}]}W(t),
\eqno{(112)}
$$
where $W(t)$ is a standard Wiener process,  $E[N^i_{[0,1]}]$ is the mean of $N^i_{[0,1]}$ under the stationary and ergodic measure, and $N^i_{[0,1]}$ is defined in (111).

{\bf Proof.} We note, that the results from \cite{BM} and from \cite{KS} implies that, by ergodic theorems,
$$
\frac{N_t}{t}\to_{t\to+\infty} \sum_{i=1}^{N}p_i^*E[N^i_{[0,1]}]
$$
or
$$
\frac{N_{nt}}{n}\to_{t\to+\infty} t\sum_{i=1}^{N}p_i^*E[N^i_{[0,1]}].
\eqno{(113)}
$$
We proceed now with similar derivations as in (78)-(82), where we have $N_t$ as non-linear Hawkes process defined in (110).  Using change of time in (82) we obtain:
$$
\hat U^*_n(N_{nt}/n)\to_{n\to+\infty}\hat\sigma^* W\Big(t\sum_{i=1}^{N}p_i^*E[N^i_{[0,1]}]\Big),
$$
or
$$
\hat U^*_n(N_{nt}/n)\to_{n\to+\infty}\hat\sigma^*\sqrt{\sum_{i=1}^{N}p_i^*E[N^i_{[0,1]}]}W(t),
\eqno{(114)}
$$
The result (112) now follows from (113)-(114).

{\bf Lemma 8 (LLN for NLRSCHPnSDO)}. The process $S_{nt}$ in (109) satisfies the following weak convergence in the Skorokhod topology:
$$
\frac{S_{nt}}{n}\to_{n\to+\infty}\hat a^*\sum_{i=1}^{N}p_i^*E[N^i_{[0,1]}]t,
\eqno{(115)}
$$
where $\hat a^*$ is defined in (92) and $E[N^i_{[0,1]}]$ in (112).

{\bf Proof.} The result (115) follows from (101)-(102), with $N(nt)$ replaced by non-linear Hawkes process $N_{nt},$ and from (113).

%\section{Numerical Examples}

\section{Conclusion} In this paper, we further studied various new Hawkes processes, namely, so-called general compound and regime-switching general compound Hawkes processes to model the price processes in the limit order books. We prove Law of Large Numbers and Functional Central Limit Theorems (FCLT) for these processes. The latter two FCLTs are applied to limit order books where we use these asymptotic methods to study the link between price volatility and order flow in our two models by studying the diffusion limits of these price processes. The volatilities of price changes are expressed in terms of parameters describing the arrival rates and price changes. 
Future work will be devoted to numerical examples associated with presented results.

\hspace{0.5cm}

\end{document}